\documentclass[useAMS,usenatbib,1column]{mn2e}

\usepackage{latexsym}
\usepackage{tabularx}
\usepackage{dcolumn}
\usepackage{amsmath,amsfonts,amssymb}
\usepackage{graphicx,epsfig}
\usepackage{color}
\usepackage{subfigure}
\usepackage{sidecap}
\usepackage{longtable}
\usepackage{float}

\newcommand{\de}{{\rm d}}
\newcommand{\bea}{\begin{eqnarray}}
\newcommand{\eea}{\end{eqnarray}}
\newcommand{\be}{\begin{equation}}
\newcommand{\ee}{\end{equation}}
\newcommand{\f}{\frac}
\newcommand{\df}{\dfrac}

\newcommand{\bc}{\begin{center}}
\newcommand{\ec}{\end{center}}

\newcommand{\T}{\rule{0pt}{3.6ex}}

\newcommand{\B}{\rule[-1.0ex]{0pt}{0pt}}

\title[Evolution of high-z LBGs]{A physical model for the redshift 
evolution of high-z Lyman-Break Galaxies}
\begin{document}

\author[C. Jose et al.]
{Charles Jose$^1$\thanks{charles@iucaa.ernet.in},
Raghunathan Srianand$^1$\thanks{anand@iucaa.ernet.in} 
and Kandaswamy Subramanian$^1$\thanks{kandu@iucaa.ernet.in}\\ 
$^1$IUCAA,Post Bag 4, Pune University Campus, Ganeshkhind, Pune 411007, India\\
}
\maketitle

\begin{abstract}
We present a galaxy formation model to understand the evolution of 
stellar mass ($M_\ast$) - UV luminosity relations, stellar mass functions and specific star formation 
rate (sSFR) of Lyman Break Galaxies (LBGs) along with their UV luminosity functions in the 
redshift range $3 \leq z \leq 8$. 
Our models assume a physically motivated form for star formation in galaxies 
and model parameters are calibrated by fitting the observed UV luminosity functions (LFs) 
of LBGs. 
We find the fraction of baryons that gets converted into stars remains nearly constant for 
$z\geq 4$ but shows an increase for $z < 4$. However, the rate of converting baryons 
into stars does not evolve significantly in the redshift 
range $3 \leq z \leq 8$. 
Our model further successfully explains the $M_\ast$ - UV luminosity 
($M_{AB}$) correlations of LBGs. 
While our model predictions of stellar mass functions compare well with the inferred data from 
observations at the low mass end, we need to invoke the Eddington bias to fit 
the high mass end. 
At any given redshift, we find the sSFR to be constant over the stellar mass range 
$5 \times 10^8 -5 \times 10^{9} M_\odot$ and the redshift evolution of sSFR is well 
approximated by a form $(1+z)^{2.4}$ for $3\leq z \leq 8$ which is consistent 
with observations. 
Thus we find that dark matter halo build up in the $\Lambda$CDM model is sufficient to explain the evolution 
of UV LFs of LBGs along with their $M_\ast$ - $M_{AB}$ relations, the stellar mass functions 
and the sSFR for $3 \leq z \leq 8$. 
\end{abstract}


\section{Introduction}

Understanding the formation and evolution of Lyman Break Galaxies (LBGs) 
in the frame work of $\Lambda$CDM cosmology is currently a very  
active area of research in extragalactic astronomy. 
In particular understanding how various physical processes affect the star 
formation in individual galaxies hosted by different dark matter halos is very 
important for understanding galaxy evolution over cosmic time. 
Additionally star formation activity in high redshift galaxies influences 
physical conditions in the intergalactic medium through ionization, mechanical and 
chemical feedback. 
In principle various parameters of the model can be constrained by direct 
observables like luminosity function and clustering of LBGs and Lyman-$\alpha$ emitters 
(LAEs) measured at different redshifts. Models constrained by such observations 
can also predict several quantities derived from observations 
like star formation rate (SFR), sSFR, stellar mass functions etc.

On the observational side, the past decade has witnessed  an ever growing 
data based on UV, optical and near Infrared deep images  
and high resolution spectra of high redshift galaxies.  
The Lyman break color selection technique  
\citep{madau_96, steidel_96_1,steidel_98,steidel_98_1} enabled the 
detection of a substantial number of faint high redshift galaxies from various 
deep field surveys 
\citep{steidel_LBG_99,giavalisco_GOODS_04,ouchi_LBG_04,beckwith_hudf,
grogin_CANDELS, illingworth_hxdf} which has 
resulted in estimates of UV luminosity functions (LF) 
of the LBGs up to $z\sim 10$ 
\citep{bouwens_07_LF_z46,bouwens_07_LF_z710,reddy_08_LF,
mclure_hudf12_12,schenker_uvlfz78_hudf12_13, oesch_hudf12_13,
lorenzoni_uvlfz79_candels_13} and also their clustering 
up to $z=5$ \citep{giavalisco_dickinson_01,ouchi_04_acf,
ouchi_hamana_05_acf,kashikawa_06_acf, hildebrandt_09_acf, savoy_11_acf, 
bielby_11_acf}. 
Further, recent advances in multi-band deep field observations 
\citep{giavalisco_GOODS_04, bouwens_color_10,bouwens_color_10_1,
bouwens_color_11,windhorst_color_11,grogin_color_11,koekemoer_color_11} 
have allowed one to infer various intrinsic quantities of 
LBGs including their age, dust content, $M_\ast$ and sSFR 
using Spectral Energy distribution (SED) 
fitting analysis \citep{stark_ellis_09_z46,
gonzalez_labbe_11, reddy_pettini_12, wilkins_gonzalez_12, gonzalez_bouwens_12, 
bouwens_illingworth_12}. Such a wealth of observations combined with theoretical models 
of galaxy formation provide a unique opportunity to probe the connection between 
early galaxies and dark matter halos hosting them. 

We have been constructing physically motivated models to explain the LFs 
and clustering of LBGs and LAEs \citep{samui_07, samui_09_lae,charles_12_clustering,
charles_13_lae_clustering} in the framework of $\Lambda$CDM cosmology. 
In our models, we assume that galaxies are formed in dark 
matter halos that sustain continuous star formation spreading over few dynamical time scales 
of the halo. The SFR is then combined with the dark matter halo formation rate to obtain 
the UV LF of high redshift LBGs. We have constrained 
the parameters related to star formation by comparing 
the model predictions of UV LFs of LBGs with observed data. 
This simple approach explains 
the connection between LBGs and LAEs at $z \geq 3$ and also with 
the dark matter halos hosting them \citep{charles_12_clustering,charles_13_lae_clustering}. 
Our model is also used to understand the galactic outflows \citep{samui_08} and 
to place an independent constraint on neutrino mass \citep{charles_11}. 

While the UV luminosity of high-z galaxies depends on the instantaneous SFR, 
the other derived quantities depend on the entire star formation history. 
Therefore, 
it is interesting to see how our models, constrained only by the UV luminosity 
functions at different $z$, predict parameters related to high-z galaxies, 
derived using SED fitting analysis.
In this paper, we investigate futher to see whether our model of 
galaxy formation with minimal assumptions can capture the basic trends 
derived from SED fitting analysis. 
We show that, within the uncertainties in the quantities derived using SED fitting 
analysis, our models successfully explain various known trends 
such as the $M_\ast-M_{AB}$ correlations, 
the evolution of sSFR and the stellar mass functions of LBGs at $z \geq 3$.  

The organization of this paper is as follows. In the next section we describe 
our basic star formation model. In section 3 we calibrate our model 
parameters using the recently updated UV LF of LBGs. In section 4 we further predict the 
$M_\ast-M_{AB}$ correlations, sSFR and stellar mass functions, compare these 
quantities with observations and discuss the implications. 
We present our conclusions in the final section. 
For all calculations we adopt a flat $\Lambda$CDM universe with cosmological 
parameters consistent with 7 year Wilkinson Microwave Anisotropy Probe (WMAP7) 
observations \citep{larson_11_wmap7} with $\Omega_m=0.27$, 
$\Omega_\Lambda=0.73$, $\Omega_b=0.045$, $h=0.71$, $n_s = 0.963$ and 
$\sigma_8 = 0.801 h^{-1}$Mpc. Here $\Omega_i$ is the background density of any 
species 'i' in units of critical density $\rho_{cric}$. The Hubble constant is 
$H_0 = 100 h$ km s$^{-1}$ Mpc$^{-1}$

\section{The Star formation model} 
\label{sec:sfr}
In \cite{samui_07} (hereafter SSS07) we have developed a physically 
motivated model to explain the LFs of high redshift LBGs. 
The crucial component of this model (see SSS07 for more details) is that 
the SFR ($\dot M_{SF}$) in a dark matter halo of mass $M$ collapsed at 
redshift $z_c$ and observed at redshift $z$ and is given by 
\citep{gnedin_96, chiu_00,choudhury_02}
\bea
\label{eqn:sfr}
\dot M_{SF}(M,z,z_c) &=& f_\ast \left(\frac{\Omega_b}{\Omega_m} M \right) 
                \frac{T}{\kappa^2 t^2_{d(z_c)}}  \\ \nonumber
  && \times \exp\left[-\frac{T}
                {\kappa t_{d(z_c)}}\right]. 
\eea
Here, $f_\ast$ is the fraction of baryons converted into stars 
over the life time of a galaxy and $\kappa$ fixes the duration of this  
star formation activity. 
Also $T =t(z)-t(z_c)$, with $t(z)$ being the age of the universe at redshift $z$, 
is the age of the galaxy at $z$ is  $t(z)-t(z_c)$. 
Further, $t_{d}(z_c)$ is the dynamical time scale of a halo collapsing 
at $z_c$ (see Eq.~(3) of SSS07).  

Such a functional form for the SFR 
in our model could be reasonable in the following way.
The SFR in a halo is proportional 
to the available cold gas. The gas in a halo is heated to the virial 
temperature of the halo at the time of it's collapse. Stars are formed 
after the cooling of this gas by radiative recombination. Thus, as the the 
available cold gas increases with time, the SFR in that halo also increases. 
The linear increase in the SFR captures this phase of star formation. 
Eventually, the amount of cold gas (and hence the SFR) in a halo decreases 
because the gas gets locked into the previous stars or due to supernova feedback 
of earlier star formation. This could give rise to the exponential 
decrease in SFR. As discussed above we have shown that the star formation 
history given by Eq.~\ref{eqn:sfr}, reproduces various observed properties of 
LBGs \citep[See also,][]{samui_14}. 

The star formation in a dark matter halo is affected by various feedback processes. 
In our models we include feedback processes related to the cooling 
efficiency of the gas and active galactic nuclei (AGN) activity.  
We assume that the gas can cool (due to recombination line 
cooling from hydrogen and helium) and collapse to form 
stars in halos with virial temperatures ($T_{vir}$) greater than $10^4$ K.
However, after reionization, the photo ionization of the IGM increases 
the temperature of the gas 
thereby increasing the Jean's mass for collapse. In such regions, we incorporate 
the radiative feedback by complete suppression of star formation in halos 
with circular velocity $v_c \leqslant 35$ km s$^{-1}$ and no suppression 
in halos with circular velocity $v_c \geq 95$ km s$^{-1}$. Further, for halos 
with intermediate circular velocities 
($35$ km s$^{-1}$ $\leq v_c \leq 95$ km s$^{-1}$) 
a linear fit from $0$ to $1$ is taken as the suppression in star 
formation \citep{bromm_02,benson_02,dijkstra_04}.  
Additionally, in our models the AGN feedback suppresses the star formation in 
high mass halos by a factor $[1+(M/M_{agn})]^{-1/2}$ \citep{charles_12_clustering}, 
where the characteristic mass scale $M_{agn}$, is believed to be 
$\sim 10^{12} M_\odot$ \citep{bower_06,best_06}. 

Since reionization leads to a feed back on the star formation, 
one should model both the star formation 
and reionization simultaneously in a self consistent manner. 
However, comparison of various predictions of our models with
observations shows that the exact nature and redshift of reionization 
does not change these predictions in the redshift range currently probed 
by observations, provided the redshift reionization, $z_{re}$, is larger 
than about $8$. Therefore, for applying radiative feedback, we assume a sudden 
reionization of the Universe at $z_{re} = 10$, 
consistent with the WMAP7 observations \citep{larson_11_wmap7}.

In summary reionization and AGN feedbacks suppresses the SFR in a given halo of mass 
$M$ by a factor 
\be
F_s(M) =\begin{cases}
   \df{1}{\left(1+\f{M}{M_{agn}}\right)^{\f{1}{2}}}         & \text{$v_c(M) > v^{m}_c $}\\
   \df{1}{\left(1+\f{M}{M_{agn}}\right)^{\f{1}{2}}}\df{v^{m}_c-v_c(M)}{v^{m}_c-35} & \text{$ v^{m}_c \geq v_c(M)>35$} \\
   0.                                         &\text{$ v_c(M) < 35$}
  \end{cases}
\ee
where $v_c$ is in units of km s$^{-1}$. 
so that the final star formation rate is given by $\dot M_{SF}(M) \times F_s(M)$. 
In our fiducial model,  as discussed above, we have adopted $v^{m}_c$ to 
be $95$ km s$^{-1}$ as suggested by \cite{bromm_02}. 

\begin{figure*}
\includegraphics[trim=1.5cm 7.0cm 1.5cm 0.0cm, clip=true, width =15.0cm, height=6.0cm, angle=0]
{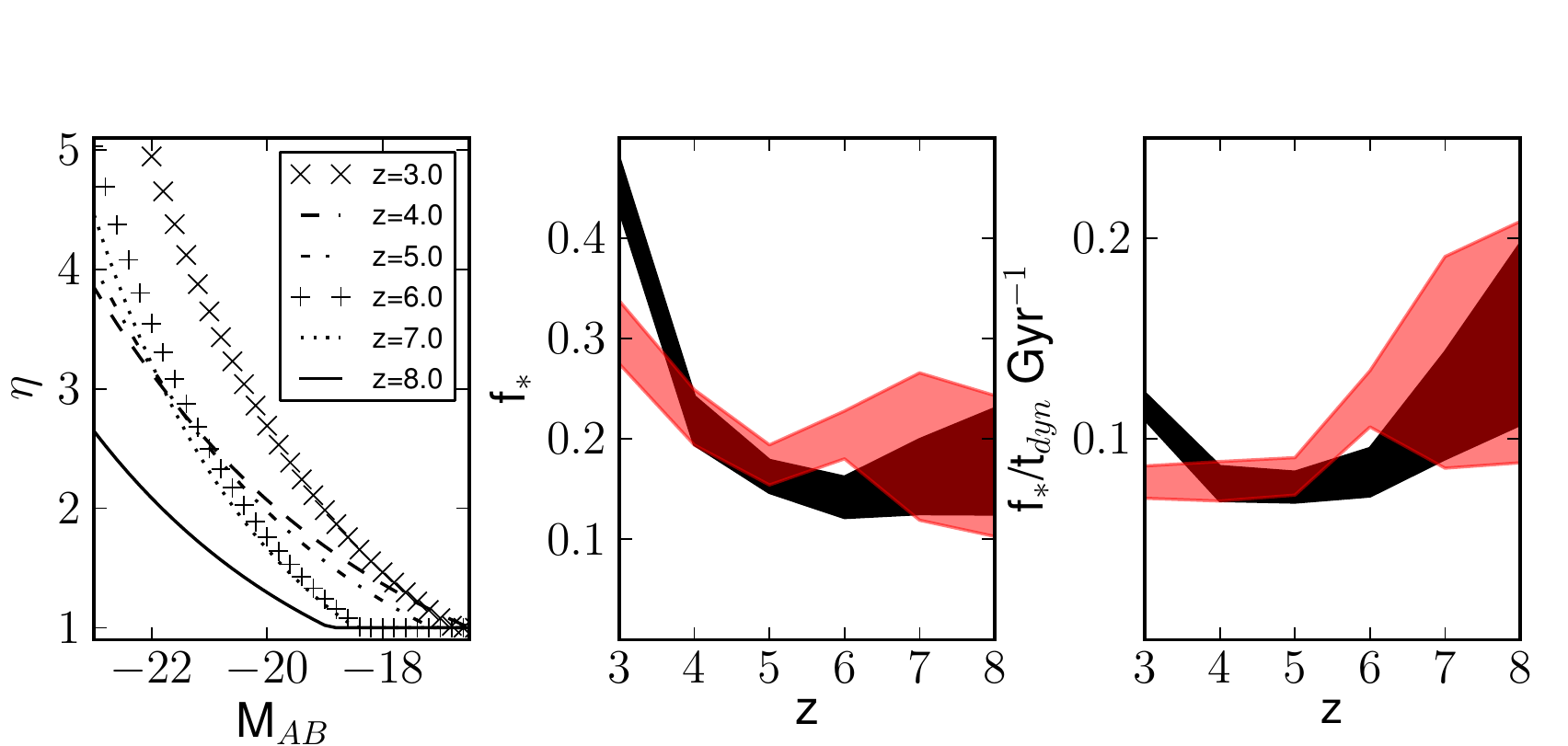} 
\caption{Left: The best fit dust correction parameter 
given by \protect \cite{bouwens_dust_09,bouwens_illingworth_12} as a 
function of the absolute AB magnitude for redshifts $3-7$. 
Middle and right panels respectively show the 1$\sigma$ range 
of $f_\ast$ and $f_\ast/t_{d}$ as a function of redshift, obtained by fitting 
our model predictions of UV LF with the observed data. 
Here the shaded black (dark) region and shaded 
red (grey) region are for our fiducial models A and B (see section 3) 
respectively. 
}
\label{fig:fstar_eta}
\end{figure*}

Given the SFR from Eq.~(\ref{eqn:sfr}) one can obtain the total $M_\ast$ 
of a galaxy at any redshift $z$ by integrating 
SFR from it's collapse redshift 
$z_c$ to the observed redshift $z$. Thus the total $M_\ast$ in a galaxy 
that has formed at $z_c$ and observed at $z$ is given by 
\be
M_\ast(M,T) = f_\ast M \f{\Omega_b}{\Omega_m} 
                      \left[1-e^{\f{-T}{\kappa t_{d}}}
                      \left(1+\f{T}{\kappa t_{d}} \right) \right] \times F_s(M). 
\label{eqn:mstar}
\ee
The total $M_\ast$ of a galaxy 
at any redshift $z$, is a function of the mass $M$ of the host dark matter halo, 
it's age and the efficiency parameter $f_\ast$. 
From the Eq.~(\ref{eqn:mstar}) it is clear that galaxies of different age 
(forming at different redshifts), hosted by dark matter halos of different  
masses can have the same $M_\ast$ at any given time of observation. 
This, as we see in next section, naturally  
introduces a scatter in various scaling relations involving $M_\ast$. 

Another quantity of great interest is the specific star formation rate (sSFR), 
which is the instantaneous star formation rate in a galaxy per unit stellar 
mass in it or simply the SFR divided by the $M_\ast$. Using Eq.~(\ref{eqn:sfr}) 
and Eq.~(\ref{eqn:mstar}) we can compute the sSFR of a galaxy at $z$, which has 
formed at $z_c$ as  
\be
sSFR(T(z,z_c))=  \df{T/(\kappa t_{d})^2} { \left[\exp\left(\f{T}{\kappa t_{d}}\right)- 
               \left(1+\f{T}{\kappa t_{d}} \right) \right] }.
\label{eqn:ssfr}
\ee
Thus in our model sSFR of a galaxy is not a function of $f_\ast$ or halo mass 
but depends only on it's age and the duration of star formation. 
Therefore all the galaxies forming 
at the same redshift will have the same sSFR at a given redshift of observation, 
independent of their halo mass. 
However, galaxies forming at two different redshifts can have very different 
sSFR at a given redshift of observation. 
Further,  sSFR decreases with the age of the galaxy, 
thereby younger galaxies forming close to the redshift of 
observation tend to posses a higher sSFR.

In order to make predictions of various derived quantities like stellar 
mass functions and average sSFR, we need to calibrate our model parameters, 
especially $f_\ast$. To do this we compare our model predictions 
of UV LFs of LBGs with their observed LFs over wide $z$ range. 
We discuss this in the next section. 

\section{Calibrating the model parameters using UV LFs of LBGs}
\label{sec:uvlf}
\begin{figure*}
\includegraphics[trim=0cm 5cm 0cm 9.5cm, clip=true, width =15.0cm, height=9.0cm, angle=0]
{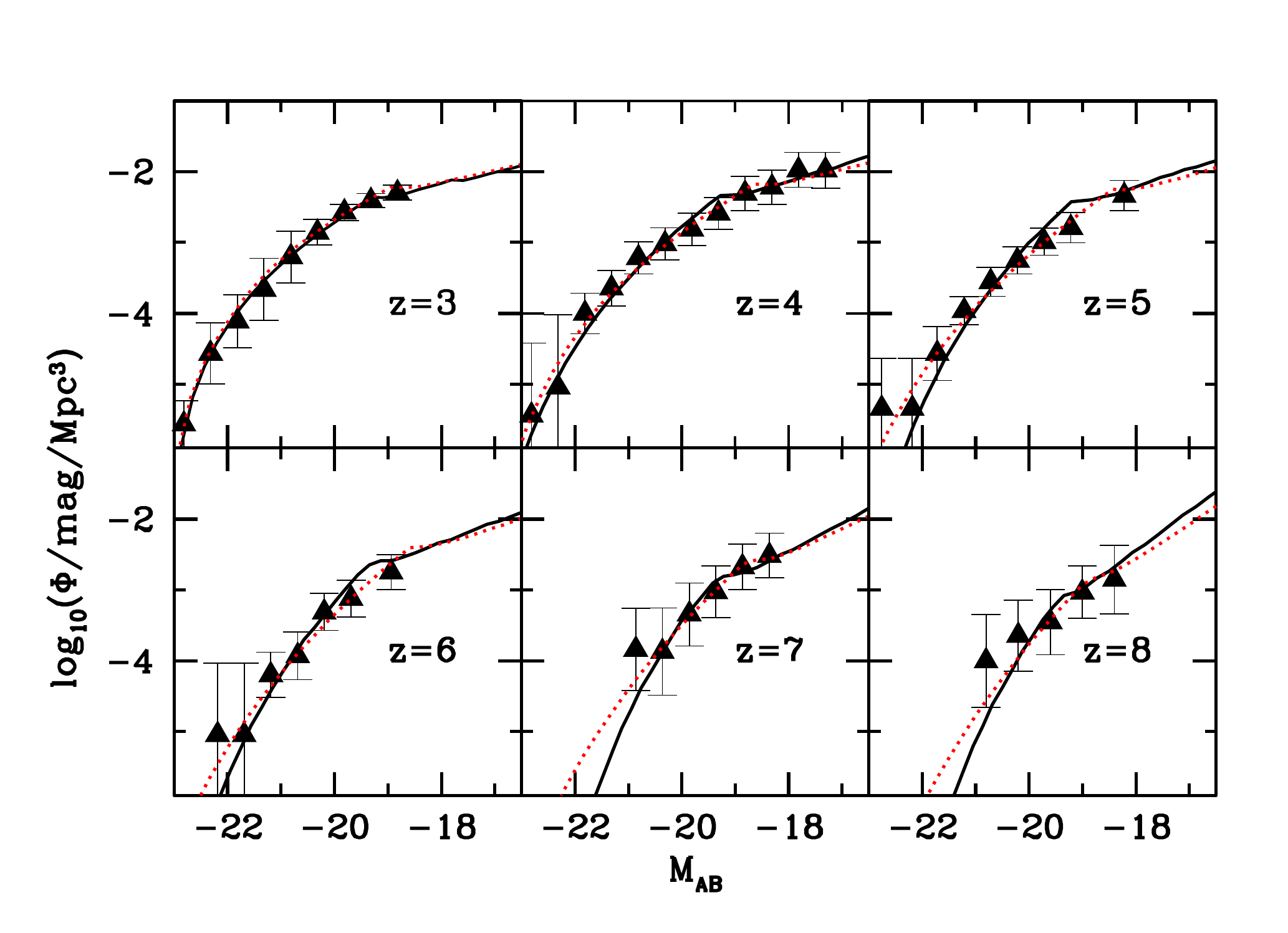} 
\caption{Comparison of observed UV LF of LBGs at different redshifts 
with our best fit model predictions. The observed data points and error 
bars are from \protect \cite{reddy_08_LF} (for $z=3$) and  \protect 
\cite{bouwens_07_LF_z46} (for $z=4-7$). 
We take cosmic variance into account by adding an uncertainty of 14 \% into 
Poisson error in quadrature for $z\geq 4$ \citep{bouwens_07_LF_z46}. 
The solid black lines corresponds 
to model A where we apply luminosity dependent dust correction and 
red dotted lines corresponds to model B where luminosity independent 
dust correction is applied. 
}
\label{fig:lfz}
\end{figure*}

As in our previous work we constrain our model parameters by fitting the 
observed UV LFs of LBGs. 
Here, we briefly describe how to compute the LFs in our models using the 
SFR given in  Eq.~(\ref{eqn:sfr}). We first note that the halo mass and 
time of collapse are input parameters for Eq.~(\ref{eqn:sfr}). 
For any observed z, we compute the SFR for the whole range of halo 
masses that collapsed at different collapse redshifts $z_c > z$. 
The number density of such halos of mass M formed at $z_c$ and observed at 
$z$ is calculated from the redshift derivative of halo mass functions (see below). 
During the period from $z_c$ to $z$, the SFR(t) in a such halo of mass M 
is given by Eq.~(\ref{eqn:sfr}) which is then used to compute the UV luminosity 
of LBGs. 

For computing galaxy luminosities, we assume that stars formed with a 
Salpeter IMF in the mass range $0.1 - 100 ~M_\odot$. This is very much similar 
to what is used by the observers to interpret their data. 
We use the population synthesis code {\sc Starburst99} \citep{starburst_99} 
to obtain the rest frame luminosity ($l_{1500}$) at 1500 \AA, as a function of time 
of a galaxy undergoing a burst of star formation.
The luminosity of a galaxy ($L_{1500}$) is obtained by convolving the SFR given by 
Eq.~(\ref{eqn:sfr}) with this burst luminosity at 1500 \AA, 
\be
L_{1500}(T) = \int_T^0 \dot M_{SF} (T-\tau) l_{1500}(\tau)d\tau 
\label{eqn:L}.
\ee

The observed luminosity is reduced to a fraction, $1/\eta$, of $L_{1500}$ due to 
the dust absorption in the galaxy, thus the observed luminosity 
is $L_{1500}/\eta$.
This luminosity ($L_0 = L_{1500}/\eta$) is then converted to a standard absolute AB 
magnitude $M_{AB}$, using 
\be
M_{AB} = -2.5 \log_{10}(L_0) + 52.60
\ee
where the luminosity is in units of erg s$^{-1}$ Hz$^{-1}$ \citep{oke_83}.
Knowing $M_{AB}$ of individual galaxies we then compute the 
LF, $\Phi(M_{AB}, z )$, at a given redshift $z$ using, 
\bea
\Phi(M_{AB}, z ) dM_{AB}&=& \int\limits_z^\infty dz_c  \frac{d^2n}{dMdz_c}(M(M_{AB}),z_c)
\frac{dM}{dL_{1500}} \nonumber\\   &&\times~ \frac{dL_{1500}}{dM_{AB}} ~dM_{AB}.
\eea

Here, 
\be
\f{\de^2 n}{\de M \de z_c}(M,z_c)= \f{dt}{dz_c}\f{\de \dot n}{\de M}(M,z_c)
\ee
where $\f{\de \dot n}{\de M}(M,z_c)$ is the formation rate 
of halos in the mass range $(M, M+dM)$ at redshift $z_c$. 
SSS07 found that using the time derivative of \cite{sheth_tormen_99} (hereafter ST) 
mass functions reproduces the 
observed LF of high-$z$ LBGs very well. Therefore we use $\f{\de \dot n }{\de M}(M,z_c)= 
\f{\de}{dt}\f{\de n_{ST}}{\de M}(M,z_c)$ with $\f{\de n_{ST}}{\de M}(M,z_c)$ being the ST 
mass function at $z_c$.


In our fiducial model we have chosen $\kappa = 1.0$ at all redshifts 
as it is consistent with the LF (SSS07) and the 
clustering \citep{charles_12_clustering} studies. 
Thus the characteristic time scale for star formation is 
set to the dynamical time scale of the halo. 
We also varied $M_{agn}$, the parameter controlling the AGN feedback, 
at each redshift to obtain the best fit LFs. 
The best fit $M_{agn}$ values are found to be $5.5 \times 10^{11}$ and  
$1.6 \times 10^{12}$ $M_\odot$ at redshifts $3$ and $4$ in our models. 
In addition we find that our model does not require any AGN 
feedback to reproduce the observed LF for $z=5-8$. 
This is mainly due to the lack of accurate LF measurements in 
the high luminosity end in these redshifts.
We also note that fixing AGN feedback mass scale to be $10^{12} M_\odot$ at all 
the redshifts also provide a reasonable fit to the UV LFs at those redshifts. 
Thus the evolution of AGN feedback parameter is not essential for our 
models. 

We use the dust extinction parameter, $\eta$, determined 
by a number of recent studies. These studies, using SED fitting analysis, 
derive (i) $\eta$ as function of the luminosity at a given redshift and  
(ii) the average luminosity function weighted $\eta$ for 
different redshifts \citep{stark_ellis_09_z46, reddy_steidal_09,
bouwens_dust_09, bouwens_illingworth_12, reddy_pettini_12}. 
Therefore, at any redshift we consider two fiducial models A and B; 
in model A we use the luminosity dependent $\eta$ and in model B, the 
luminosity function weighted average $\eta$. 

The luminosity dependent $\eta$ for model A is computed using the  
recent estimates of UV-continuum slope $\beta$ 
from \citet{bouwens_illingworth_12}.
The dust extinction ($\eta$) is related to UV-continuum slope through 
the IRX-$\beta$ relation 
$2.5 \log_{10}(\eta) =4.43+1.99\beta$ \citep{meurer_99}. 
Using SED fitting analysis, 
\cite{bouwens_illingworth_12} provides $\beta$ as a function of the UV magnitude 
$M_{AB}$ in the redshift range $4-7$(see Table 5 of their paper). 
In this work, we use the values of $\beta$ and $d\beta/dM_{AB}$ at $M_{AB} = -19.5$ 
given by \cite{bouwens_illingworth_12} and obtain UV-continuum slope at other 
magnitudes by assuming a linear relation between $\beta$ and $M_{AB}$. Further,  
for redshifts $3$ and $8$, where observational data is not available 
we use $\beta$ and $d\beta/dM_{AB}$, extrapolated from other $z$. 
The values of $\beta$ and $d\beta/dM_{AB}$  
at $M_{AB} = -19.54$ are also given in Table~\ref{tab1}.  
The luminosity dependent $\eta$ computed in this way for different $z$ is 
plotted in the left panel of Fig.~\ref{fig:fstar_eta} 
\citep[see also Fig.~14 of][]{bouwens_illingworth_12}. 
The average $\eta$ used in our model B is taken from 
\cite{bouwens_dust_09} for $z=3$ and \cite{bouwens_illingworth_12} for $z=4-7$ 
\citep[see Table 6 of][]{bouwens_illingworth_12}. At any given redshift 
this is obtained by integrating over the distribution of $\beta$ and using 
IRX-$\beta$ relationship for a given limiting magnitude.  
In our model B, we adopt the average $\eta$ for a limiting magnitude of $M_{AB}=-16$ 
and this is given in Table~\ref{tab1}. 
From Fig.~\ref{fig:fstar_eta} and Table~\ref{tab1} it is clear that 
the dust correction at $z=3$ is at least $1.7$ times higher than that at other 
redshifts. This has interesting implications to the redshift evolution of 
$f_\ast$ estimated as we see below.

We found that the LF predictions of model A compares well with observations 
when suppression due to feedback as described above is 
applied over $35 \leq v_c$ (km s$^{-1}$)$ \leq 110$ (or $v^{m}_c$ is taken 
to be 110 km s$^{-1}$). 
The radiative feedback is not known to operate in  
halos with circular velocity greater than $95$ km s$^{-1}$ \citep{bromm_02}. 
Therefore, the requirement for suppression of SFR in halos with 
$v_c \geq 95$ km s$^{-1}$ could be a signature of the presence of additional 
feedback mechanisms such as supernovae feedback operating in these galaxies. 
On the other hand, for model B, we only need the fiducial feedback as given 
by \cite{bromm_02} to get a reasonable fit to the LFs of LBGs. 

Given the above model parameters the remaining crucial parameter $f_\ast$, 
the fraction of baryons being converted into stars over the life time of the 
galaxy, is fixed by fitting the observed luminosity function of LBGs using 
$\chi^2$ minimization. 
We note that if we did not have an estimate of $\eta$ as above, we would only be able 
to determine the ratio $f_\ast/\eta$ as in SSS07. 
In this manner, our physically motivated model 
finds the relationship between the halo mass (M) and the luminosity 
(L) of the galaxy it hosts. 
In Fig.\ref{fig:lfz}, we compare the 
LF predictions in the redshift range $3-8$ by both models A (in solid 
black lines) and B (in dotted red lines) with the observed data. 
The $f_\ast$ and reduced $\chi^2$ obtained by comparing 
our model predictions with the observed data at various redshifts are 
tabulated in Table~\ref{tab1}. 
From the figure, it is clear that, the number density of brightest galaxies 
($M_{AB} < -20$) as predicted by model A drops faster than that of 
model B for $z\geq 5$. However, both our models compare very well with the 
observed UV LF of LBGs given by \cite{reddy_08_LF} for $z=3$ and 
\cite{bouwens_07_LF_z46} for $z=4-8$. 

The best fit $f_\ast$ 
for model A (shown in Fig.\ref{fig:lfz}) are 
$0.34, 0.24, 0.19, 0.23, 0.20$ and $0.19$
respectively for $z=3-8$. For model B, where we use the 
luminosity independent $\eta$ the corresponding values of $f_\ast$ are 
$0.46, 0.23, 0.17, 0.16, 0.17$ and $0.20$.  
In the middle panel of Fig.~\ref{fig:fstar_eta} we show the  
1$\sigma$ range (corresponding to $\Delta \chi^2 = 1$) of $f_\ast$  
as a function of redshift for 
model A in shaded black (dark) region and for model B in shaded red (grey) region. 
The best fit values of $f_\ast$ are also tabulated in Table~\ref{tab1}. 

From Table~\ref{tab1} and Fig.~\ref{fig:fstar_eta}, it is clear that, $f_\ast$, 
the fraction of baryons being converted into stars, 
shows only a minor evolution from $z=8$ to 
$4$ in both our models. On the other hand, at $z=3$, $f_\ast$ is roughly $1.7$  
times larger than that at higher redshifts. To understand this, we note that 
our models predict a nearly constant light to mass ratio ($f_\ast/\eta$), at 
all redshifts. But the dust extinction, $\eta$, 
at $z=3$ is roughly $2$ times the dust extinction in the redshift range $4-8$ 
(see left panel of Fig.~\ref{fig:fstar_eta} and  Table~\ref{tab1}). 
Thus $f_\ast$ is significantly higher at $z=3$ compared to 
best fit $f_\ast$ at other redshifts. 
This implies that the fraction of baryons converted into stars shows a modest evolution 
in the redshift range $8-4$, but subsequently increases faster from $z=4$ to $3$.

On the other hand, SFR in galaxies at any redshift roughly scales as $f_\ast$/$t_{d}$, 
where $t_{d}$ is the dynamical time scale at that redshift.   
We have plotted the 1$\sigma$ range of $f_\ast$/$t_{d}$ in the right panel 
of Fig.~\ref{fig:fstar_eta} as a function of redshift for model A in 
shaded black region (dark) and for model B in shaded red region (grey). 
From the figure it is clear that $f_\ast$/$t_{d}$ in dark matter halos shows no 
significant evolution from $z=8$ to $3$ which suggest that SFR of galaxies do not change 
considerably in this time period. This in turn implies that the evolution of dark matter 
halo mass function can account for the redshift evolution of global star formation rate density 
of LBGs from $z=8$ to $3$  without requiring any redshift dependent efficiency of 
converting baryons into stars which 
is consistent with previous studies \citep{tacchella_12,behroozi_13B}. 

\begin{table*}
\tabcolsep 6.0pt
\begin{tabular}{cccccccccccc}\hline
$z$    &\multicolumn{6}{c}{Model A}  &\multicolumn{5}{c}{Model B} \T\T \\ 
~          &$\beta$  &$d\beta/dM_{AB}$  &$f_\ast$  &$\chi^2_\nu$   &sSFR  &sSFR 
                     &$\eta$  &$f_\ast$ &$\chi^2_\nu$   &sSFR  &sSFR \B\B\B   \\
(1)   &(2) &(3) &(4) &(5)  &(6)  &(7) &(8) &(9) &(10) &(11) &(12)  \B\B\B \\
\hline  
\T \B 3    &-1.769  &-0.166      &0.34  &0.56  &0.96  &1.55   &3.7  &0.46  &0.87  &0.92  &1.44 \\
\T \B 4    &-1.886  &-0.113      &0.24  &1.01  &1.89  &1.84   &2.2  &0.23  &0.76  &1.80  &1.75 \\
\T \B 5    &-1.920  &-0.130      &0.19  &1.60  &3.20  &2.68   &2.1  &0.17  &1.07  &3.16  &2.81\\
\T \B 6    &-2.013  &-0.191      &0.23  &0.79  &4.48  &4.10   &1.6  &0.16  &0.87  &4.23  &4.22\\
\T \B 7    &-2.040  &-0.180      &0.20  &0.50  &5.60  &5.72   &1.5  &0.17  &0.36  &4.36  &5.76\\
\T \B 8    &-2.150  &-0.130      &0.19  &0.72  &6.40  &7.31   &1.2  &0.20  &1.01  &3.60  &7.31\\
\hline
\end{tabular}
\caption{Parameters and predictions of our models. 
Column (1) redshift; 
columns (2) and (3) the UV-continuum slope $\beta$ and $d\beta/dM_{AB}$ at 
$M_{AB} = -19.5$ which is used to compute the dust extinction for model A; 
columns (4) and (9) $f_\ast$ for models A and B; 
columns (5) and (10) reduced $\chi^2$ obtained for model A and B; 
columns (6) and (11) sSFR at $5 \times 10^{10} M_\odot$ predicted by model A and B;
columns (7) and (12) sSFR at $10^{10} M_\odot$ predicted by model A and B;
column (8) average dust extinction used in model B; 
}
\label{tab1}
\end{table*}

\section{The star formation history of LBGs}
In this section, using $f_\ast$ values given in Table ~\ref{tab1} as a function 
of $z$, we compute $M_\ast$ and sSFR of LBGs. 
Unlike the LFs, these parameters are derived from the observations 
using SED fitting technique that assumes IMF, star formation history, 
dust and metallicity. It should be noted that there are 
strong degenerecies between these parameters \citep{bouwens_illingworth_12,mitchell_2013, 
finlator_2007,schaerer_13}, especially when small number of photo metric points 
are used in the SED fitting.

While we use an IMF similar to what is used in SED fitting analysis, the 
star formation history in our models is different from those typically 
used in the SED fitting analysis. Further, 
when modelling stellar mass functions we incorporate systematic 
uncertainties in the $M_\ast$ due to the typical assumptions 
about dust, metallicity etc in the SED fitting analysis  \citep{mitchell_2013}. 
A systematic exploration of 
uncertainties in the dust extinction and derived $M_\ast$ due to 
the effect of various degenerate parameters on SEDs of galaxies is outside 
the scope of the present work. 
Therefore, while comparing the model 
predictions with those derived from SED fitting methods, our main 
emphasis is to compare the overall trends and not to get exact matching of 
absolute values.

\subsection{The $M_{AB}$ - Stellar Mass relationship.}
\begin{figure}
\includegraphics[trim=0cm 4cm 0cm 4cm, clip=true, width =7.5cm, height=7.0cm, angle=0]
{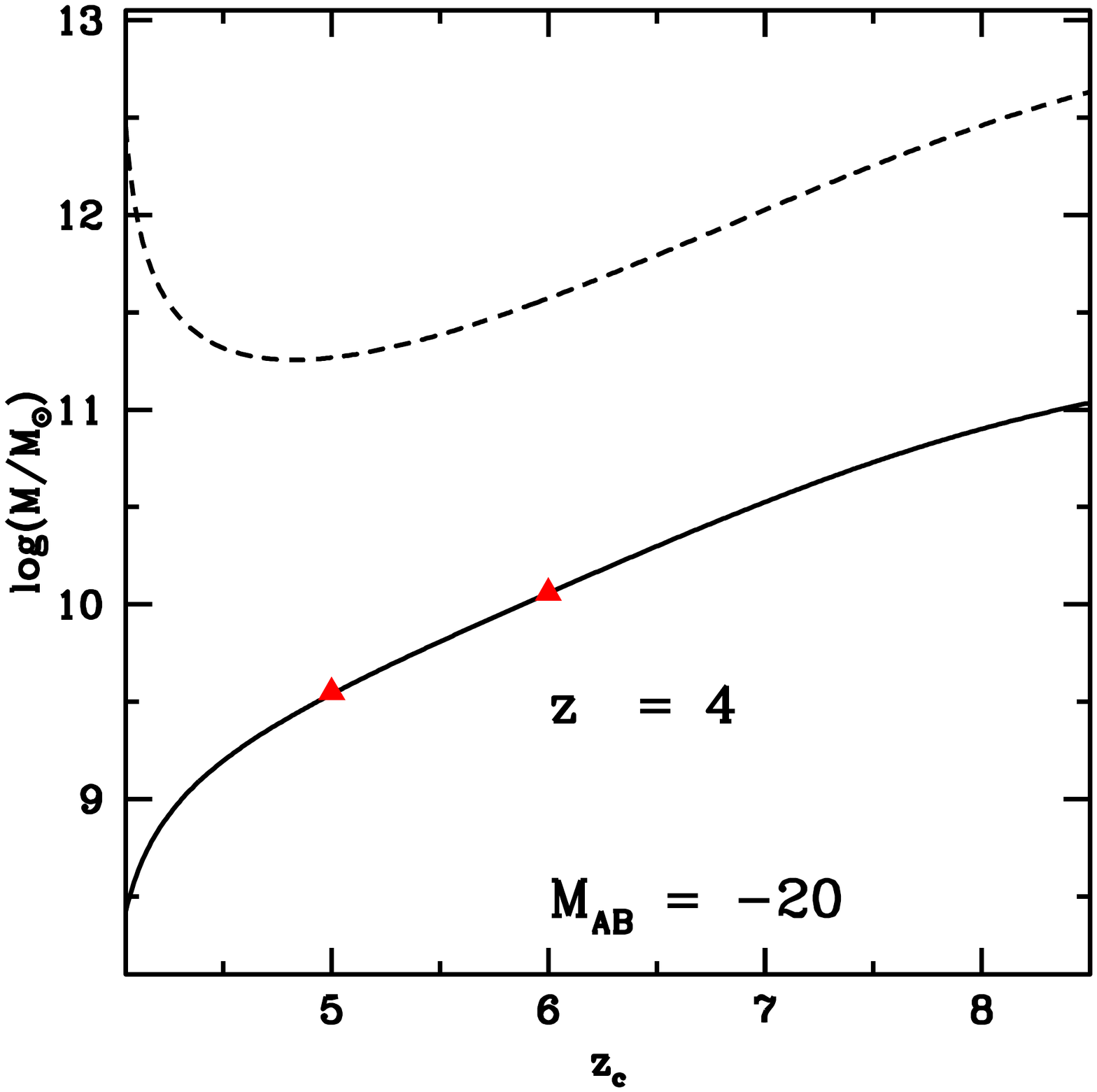} 
\caption{The stellar (solid line) and dark matter (dashed line) 
halo masses of galaxies that shine with 
$M_{AB}=-20$ at $z=4$ as a function of their formation 
(collapse) redshift $z_c$. 
The red triangles show the $M_\ast$ in galaxies formed at $z=5$ and $6$ 
and are respectively $3.4\times10^9 M_\odot$ and $1.1\times10^{10} M_\odot$. 
}
\label{fig:mstar_zc}
\end{figure}

It is clear from Eq.~\ref{eqn:mstar} that the $M_\ast$ of galaxies 
in our models are directly proportional to the halo mass. 
Since the average UV luminosity of galaxies also increases with their mass, we expect 
correlation between $M_\ast$ of galaxies and their luminosities. 
Further, galaxies of a given absolute magnitude are hosted 
by dark matter halos of different masses collapsed at different redshifts. 
At the time of observations these galaxies can have 
various $M_\ast$ depending on their halo 
mass and collapse redshift (or age). Therefore, galaxies of a given 
absolute magnitude can have a scatter in their $M_\ast$. 
This is apparent from Fig.~\ref{fig:mstar_zc}, where we have plotted the 
$M_\ast$ and halo mass of galaxies which shine with an absolute 
magnitude $M_{AB} = -20$ at $z=4$ against their redshift of formation 
($z_c$). The solid black line shows the $M_\ast$ and the dashed black 
line shows the halo masses of these galaxies as a function of $z_c$. 
The figure clearly shows that, 
in our models, galaxies of a given $M_{AB}$ observed at any $z$, 
will have larger $M_\ast$ if they collapsed earlier (larger $z_c$).
For example, one can see from Fig.~\ref{fig:mstar_zc} that, 
for observed $M_{AB}=-20$ at $z=3$, the $M_\ast$ 
of galaxies that are formed at $z_c=5$ is $3.4\times10^9 M_\odot$ 
whereas the $M_\ast$ of a galaxy with $z_c=6$ is 
$1.1\times10^{10} M_\odot$. These masses are shown in red triangles. 

\begin{figure*}
\includegraphics[trim=0cm 0.0cm 0cm 0.0cm, clip=true, width =15.0cm, height=5.0cm, angle=0]
{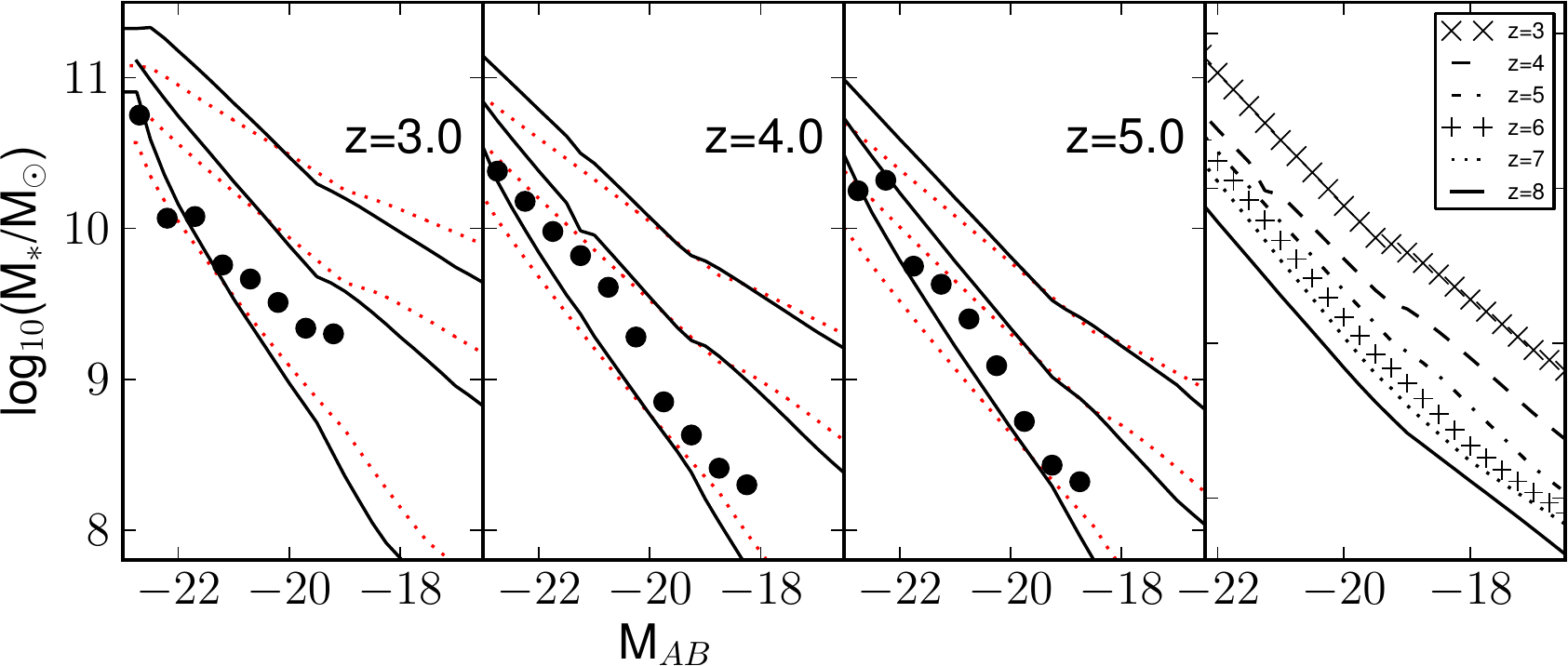} 
\caption{First three panels from the left show the mean $M_\ast$ of LBGs as a function of their 
absolute AB magnitudes for three independent redshifts along with the $1\sigma$ (68.4\%) 
dispersion in $M_\ast$ around it's mean value. The solid black curves are for model 
A and red dotted curves are for model B. The observed data points are from 
\protect \cite{reddy_pettini_12} at $z=3$ and from 
\protect \cite{lee_ferguson_12} at $z= 4$ and $5$. The right most panel shows the 
mean $M_\ast$ of LBGs as a function of their $M_{AB}$ at various redshifts.   
}
\label{fig:mag_mstar}
\end{figure*}

We compute the mean $M_\ast$ of galaxies of a given $M_{AB}$ at any $z$ as 
\be
\langle M_\ast(M_{AB},z) \rangle = 
    \df{\int_z^\infty~dz_c~M_\ast(M, z_c)~\f{d^2 n}{dM~dz_c}(M,z_c)}
       {\int_z^\infty~dz_c~\f{d^2 n}{dM~dz_c}(M,z_c)}. 
\ee
Here, $M_\ast$ is given by Eq.~\ref{eqn:mstar} and $M = M(M_{AB},z, z_c)$ 
is the mass of the dark matter halo that has 
formed at $z_c$ and hosting a galaxy which shines with brightness 
$M_{AB}$ at $z$.
In the first three panels of Fig.~\ref{fig:mag_mstar}, we show the 
predictions of mean $M_\ast$ in solid black lines for model A with luminosity 
dependent $\eta$ at various redshifts as a function of their absolute 
magnitude at 1500 \AA. 
The dotted red curves are the predictions of mean $M_\ast$ for 
model B with luminosity independent $\eta$. Because of the scatter in 
$M_\ast$ at a given magnitude, we also show in Fig.\ref{fig:mag_mstar}, 
the $1\sigma$ (68.4\%) dispersion in $M_\ast$ around it's mean value. 
The data points (solid circles) are the mean $M_\ast -M_{AB}$ 
relation derived using SEDs at $z=3$ by \cite{reddy_pettini_12} and at 
$z=4$ and $5$ by \cite{lee_ferguson_12}.

We note from Fig.\ref{fig:mag_mstar} that the predicted trend  
of $M_\ast-M_{AB}$ relations at $z=3-5$ matches with the observed data 
of \cite{reddy_pettini_12} and \cite{lee_ferguson_12} within the 
$1\sigma$ range.  
The slopes of $M_\ast-M_{AB}$ 
relation as predicted by our models are comparable to the ones given by these authors. 
For example, at $z=4$ \cite{lee_ferguson_12} fits their $M_\ast-M_{AB}$ with 
a double power with bright end slope of $-0.414$ for galaxies brighter than 
$M_{AB} = -20.45$. Our theoretical predictions follow a single power law with 
slopes $\sim -0.41$ for model A and $\sim -0.35$ for model B. 
On the other hand, \cite{gonzalez_labbe_11} derived 
a single power law slope of $-0.68$ for the $M_\ast-M_{AB}$ but with a large  
standard deviation of $0.5$ dex around the best fit $M_\ast$. Their scaling 
relations are in general consistent with the findings of 
\cite{lee_ferguson_12} and hence with our model predictions as well. 
The $M_{AB}$-$M_\ast$ relations obtained from our models are also consistent with 
many of the current observations by \cite{stark_ellis_09_z46,gonzalez_labbe_10, 
gonzalez_bouwens_12}. 

We also show in the last panel of Fig.\ref{fig:mag_mstar}, the mean $M_\ast$ of 
LBGs as a function of their $M_{AB}$ for $3 \leq z \leq 8$. 
Many of the earlier studies \citep{gonzalez_labbe_10,gonzalez_bouwens_12,lee_ferguson_12} 
find little evolution in the $M_{AB}-M_\ast$ relation with redshift. 
However, \cite{stark_schenker_13} found that the effect of nebular emission 
in the SEDs of high-z LBGs decreases the derived $M_\ast$ for $z \geq 5$. 
They showed that, this will tentatively result in the evolution of 
the $M_\ast-M_{AB}$ relations with redshift. From their analysis they find 
a decrease in the over all normalization of the $M_\ast-M_{AB}$ relation by a 
factor of $1.4-2.5$ in the redshift range $4$ to $7$. Interestingly our models 
also predict a decrease in the overall normalization of the $M_\ast-M_{AB}$ 
relation by a factor $\sim 3$ in the above mentioned range. 
\cite{stark_schenker_13} further noted that the reduction in the normalization 
of the $M_\ast-M_{AB}$ relation for $z \geq 5$ is crucial for obtaining a 
redshift evolution of the derived sSFR as predicted by many theoretical studies.

Finally, we note that the mean $M_\ast$ predicted by our models at given magnitude 
are systematically higher than the predictions of \cite{reddy_pettini_12} and 
\cite{lee_ferguson_12}.  
This could be due to the uncertainties associated with (i) the $M_\ast$ of individual 
galaxies predicted by our model using $f_\ast$ which is prone to systematic 
uncertainties discussed earlier and (ii) the $M_\ast$ derived 
by \cite{reddy_pettini_12, lee_ferguson_12} using SED fitting analysis.
The exponentially increasing and constant SFHs used respectively by 
\cite{reddy_pettini_12} and \cite{lee_ferguson_12} for SED fitting analysis can result 
in significant decrease in inferred $M_\ast$ compared to the $M_\ast$ inferred using 
SFH in our models \citep{finlator_2007,schaerer_13}.

\subsection{The stellar mass functions}
\begin{figure*}
\includegraphics[trim=0cm 5.0cm 0cm 5.0cm, clip=true, width =14.0cm, height=12cm, angle=0]
{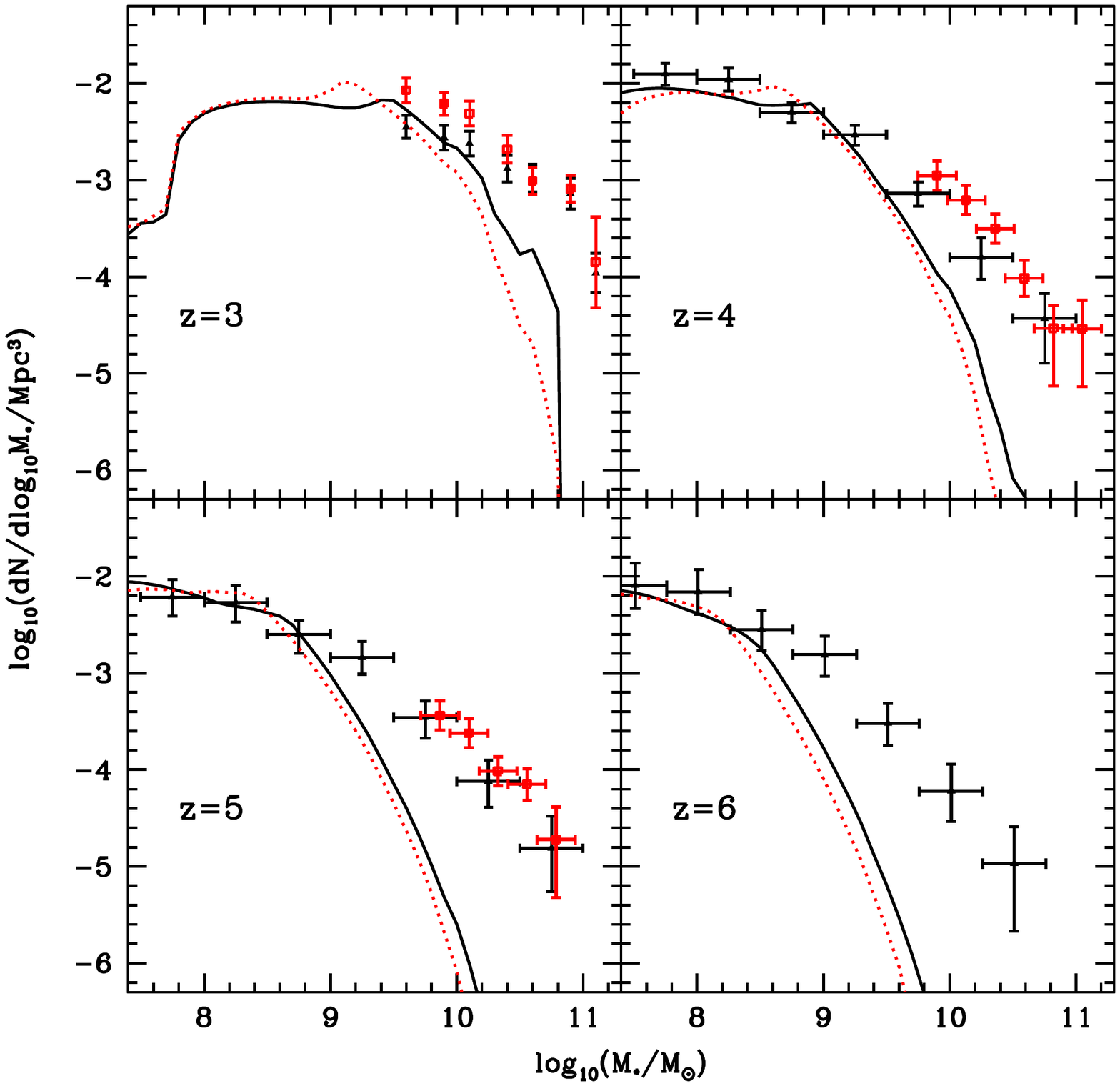} 
\caption{The stellar mass functions as predicted by our models from redshifts $3$ to $6$. 
The predictions of model A and model B are respectively shown in solid-black red-dotted lines. 
The data points and error bars are from 
(i) \protect \cite{mortlock_smf_z3} in black triangles for $z = 3-3.5$ 
and in red squares for $z=2.5-3$,  
(ii) \protect \cite{gonzalez_labbe_11} in black triangles for $z=3$ to $6$ 
and (iii) \protect \cite{lee_ferguson_12} in red squares for $z=4$ to $5$. 
}
\label{fig:smf}
\end{figure*}

The stellar mass function, $\Phi(M_\ast,z)$, is the number density of galaxies of a 
given $M_\ast$ at a given redshift, $z$. It can be computed as 
\be
\Phi(M_\ast,z)~dM_\ast = \int_z^\infty \f{d^2 n}{dM~dz_c}(M,z_c)~dz_c~
                     \f{dM}{dM_\ast}~dM_\ast. 
\ee
Here the halo mass $M = M(M_\ast,z_c)$ is related to the $M_\ast$ 
by Eq.~\ref{eqn:mstar}. We show our model predictions of stellar 
mass functions in Fig.~\ref{fig:smf} for $z=3-6$. We also show in this figure, the data of 
\cite{mortlock_smf_z3} at $z=3$ and \cite{gonzalez_labbe_11, lee_ferguson_12} 
at other redshifts. Here, our basic model predictions with luminosity dependent 
dust correction (model A) are shown in solid-black lines. The red-dotted lines show 
predictions of our model B with constant dust extinction. 
While computing the stellar mass functions at each redshift, we consider only those 
galaxies with $M_{AB} \leq -15$,  which is the luminosity selection  
criteria of \cite{gonzalez_labbe_11}. 

We firstly note that our model predictions of stellar mass functions flattens 
at low mass end rather than being a power law. This flattening is due to 
the the feedback operating in the low mass galaxies and 
the magnitude threshold criteria used for the galaxy sample. Both our models 
produce almost identical stellar mass functions with model A producing 
slightly higher number density of galaxies at the high $M_\ast$ end. 
In general our model predictions compare reasonably well with the observed 
stellar mass functions of \cite{mortlock_smf_z3} at $z=3$ for 
$M_\ast \le 10^{10} M_\odot$ and with that of \cite{gonzalez_labbe_11, lee_ferguson_12} 
for $M_\ast \le 10^9 M_\odot$ at other redshifts.  
However, we note that they systematically underpredict stellar 
mass functions typically for $M_\ast \ge 10^{10} M_\odot$. 
The discrepancy between observationally derived stellar mass functions and 
theoretical predictions increases with increasing redshift as well. 
This apparent deficiency in our models could be due to a number of reasons 
which we will discuss now.  

First, the different SFHs used for SED fitting analysis would 
give different $M_\ast$ for the galaxies observed. \citet{schaerer_13} showed that 
the constant SFH used by \cite{gonzalez_labbe_11} for their analysis systematically 
predicts a larger $M_\ast$ compared to the SFH used by our models. Earlier studies of 
\cite{finlator_dave_07} suggested that the differences between $M_\ast$ obtained 
via SEDs when one uses different SFHs are typically less than $0.3$ dex. 
Using simulations and after considering the effects of dust, age, SFH and photometric 
uncertainties in SEDs, \cite{lee_ferguson_12} found that, $M_\ast$, on average, 
can be recovered within $0.1$ dex for LBGs. \cite{debarros_12} and 
\cite{stark_schenker_13} found that not incorporating nebular emissions in 
the SEDs of high-z LBGs also systematically overpredicts $M_\ast$ for LBGs 
at high redshift. The data presented by \cite{mortlock_smf_z3, gonzalez_labbe_11, 
lee_ferguson_12} do not consider the effects of nebular emission.
All of these effects discussed above will result in a shift in 
the observed stellar mass functions to the high mass end. 

\begin{figure}
\includegraphics[trim=0cm 0.0cm 0cm 0.0cm, clip=true, width =8.0cm, height=6.5cm, angle=0]
{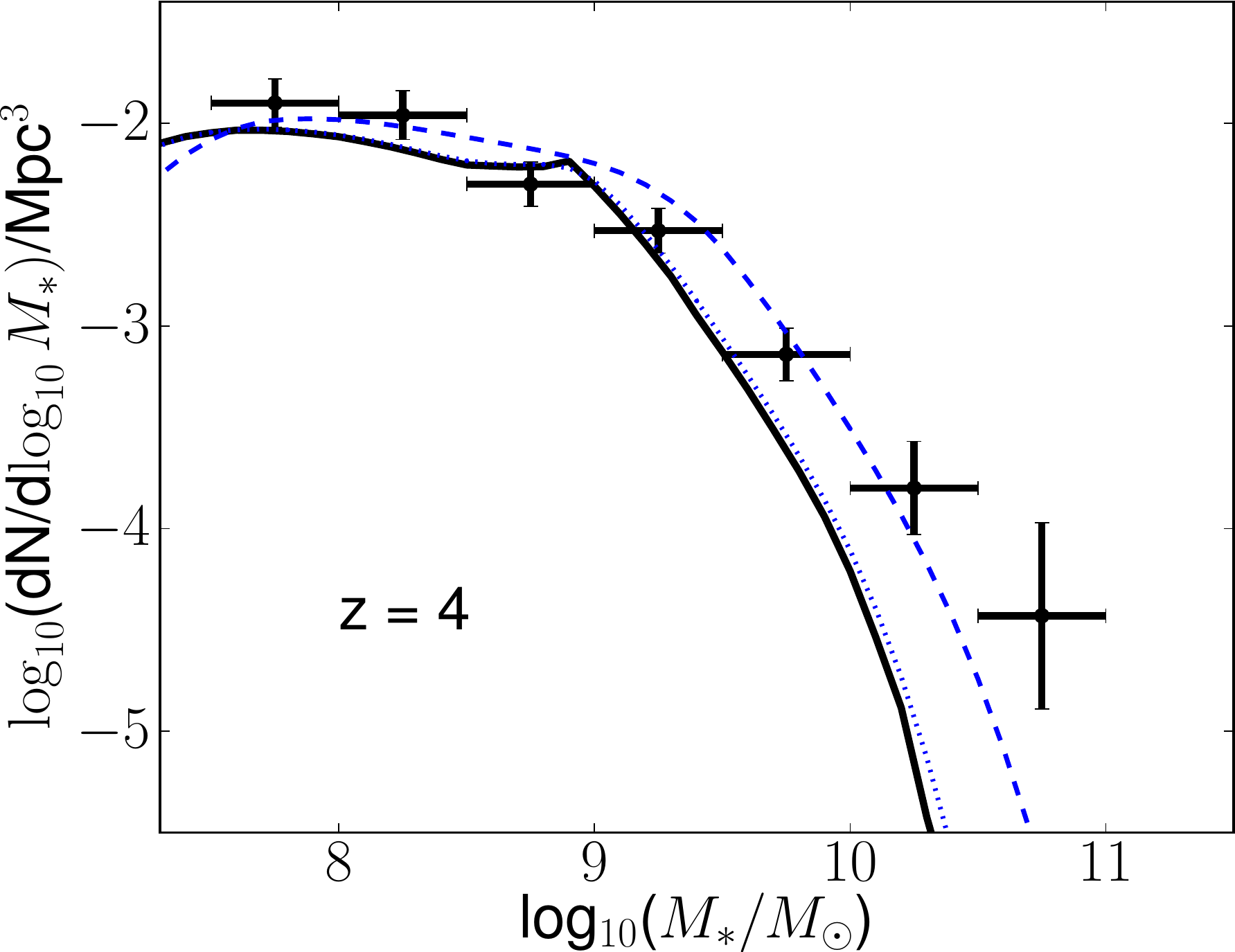} 
\caption{The effect of scatter in $M_\ast$ in our model predictions of 
stellar mass functions at $z=4$. 
The predictions of the model with luminosity dependent $\eta$ is given in solid black 
lines. The blue dotted and dashed lines are our model predictions of stellar 
mass functions which assumes a Gaussian distribution of variance $0.1$ and 
$0.3$ dex respectively in the $M_\ast$. 
}
\label{fig:smf-cor}
\end{figure}

More importantly, due to various systematic uncertainties, the SED fitting 
analysis do not recover the true $M_\ast$ of the observed galaxies accurately. 
They  rather result in a distribution of the recovered $M_\ast$ around the true 
value of $M_\ast$ of galaxies \citep{lee_ferguson_12,schaerer_13,mitchell_2013}. 
Such a scatter is evident from Fig.~5 of \cite{lee_ferguson_12} where the 
authors have considered effects of age, dust, SFH and uncertainties related 
to photometry on SEDs. \cite{schaerer_13} also show  in Fig.~4 of their paper 
that such a scatter in $M_\ast$ is expected when one uses different star 
formation models. 

In particular, \cite{mitchell_2013} have shown 
how the $M_\ast$ obtained by fitting SEDs is affected by the typical assumptions 
about SFH, metallicity, dust extinction etc in the population synthesis 
models \citep[See also, ][]{moster_13}. 
Various SED fitting studies assume fixed values of metallicity and dust 
for a given type of galaxies whereas in real galaxies these quantities vary 
stochastically.  
\cite{mitchell_2013} showed that such assumptions about metallicity and dust 
in the SED fitting analysis will result in a large scatter 
in the SED derived $M_\ast$ around the true value of $M_\ast$. 
More interestingly the scatter in the derived $M_\ast$ increases 
with increase in the true $M_\ast$ of galaxies and also with the 
redshift of observation. Thus the observationally derived stellar masses of 
high mass galaxies are much more sensitive to systematic uncertainties 
compared to that of low mass galaxies. 
For example at $z=4$ and $M_\ast \sim 10^{10} M_\odot$, the typical scatter 
in $M_\ast$ can be as large as $0.5$ dex whereas on smaller mass scales like 
$10^8 M_\odot$ the scatter in $M_\ast$ is $\sim 0.1$ dex. 
This important effect results in so called Eddington bias 
\citep{ eddington_1913,eddington_1940} in the observed stellar mass functions, 
where many low mass galaxies are up scattered to the high mass end. 

We have investigated the effect of Eddington bias in our model predictions of 
stellar mass functions, due to the scatter in the recovered $M_\ast$ 
compared to the true $M_\ast$. 
To do this we assumed that the stellar mass function at any recovered $M_\ast$ 
has contributions from all the other intrinsic $M_\ast$ because of the above mentioned scatter.
For simplicity, we have assumed that this scatter is given by a Gaussian distribution 
of variance $\sigma_{\log M_\ast}$ dex. Assuming $M'_\ast$ to be the true stellar 
mass and $M_\ast$ to be the recovered stellar mass, the modified stellar mass 
function ($\Phi_G$) after taking care of the Eddington bias can be computed from 
the actual stellar mass function ($\Phi$) as 
\bea
\Phi_G(M_\ast) =&& \int dM'_\ast \Phi (M'_\ast) \f{1}{\sqrt{2\pi\sigma^2_{\log M_\ast}}} \nonumber \\
&&~~\exp\left[-\df{1}{2}\left( \f{\log(M'_\ast) -\log(M_\ast)} {\sigma_{\log M_\ast}} \right )^2 \right]. 
\label{eq:smf-smooth}
\eea

In Fig.~\ref{fig:smf-cor}, we demonstrate the effect of Eddington bias on stellar mass 
functions. 
Here the solid black curve corresponds to predictions of model A with luminosity 
dependent dust correction ($\sigma_{\log M_\ast}= 0)$. In this figure we also show 
the modified stellar mass function $\Phi_G(M_\ast)$ for models with 
$\sigma_{\log M_\ast}= 0.1$ and $0.3$ dex in dotted and dashed blue lines.
The models with $\sigma_{\log M_\ast}= 0.1$ dex do not alter 
stellar mass functions appreciably. 
Since the scatter in derived $M_\ast$ are small for low $M_\ast$ 
we expect the low mass end of the stellar mass functions to be not 
affected by the Eddington bias. 
On the other hand, a scatter of $0.3$ dex in the $M_\ast$ considerably changes 
the slope of stellar mass functions at high mass end and better represent the 
observed data. Also, as already mentioned, the scatter in 
the $M_\ast$ derived from SEDs are much larger for high $M_\ast$.  
We further note that, the stellar mass function
has a steep fall at the high stellar mass end. 
Therefore, even a constant error across all masses will widen 
the distribution, thereby enhancing the number of inferred galaxies 
at the high stellar mass end. 
Therefore, we conclude that effect of Eddington bias is important  
while modelling high redshift LBG stellar mass functions and that 
our models are consistent with the data within the allowed Eddington bias.  

\subsection{The specific star formation rate}
\begin{figure}
\includegraphics[trim=0cm 0cm 0cm 0cm, clip=true, width =7.5cm, height=6.5cm, angle=0]
{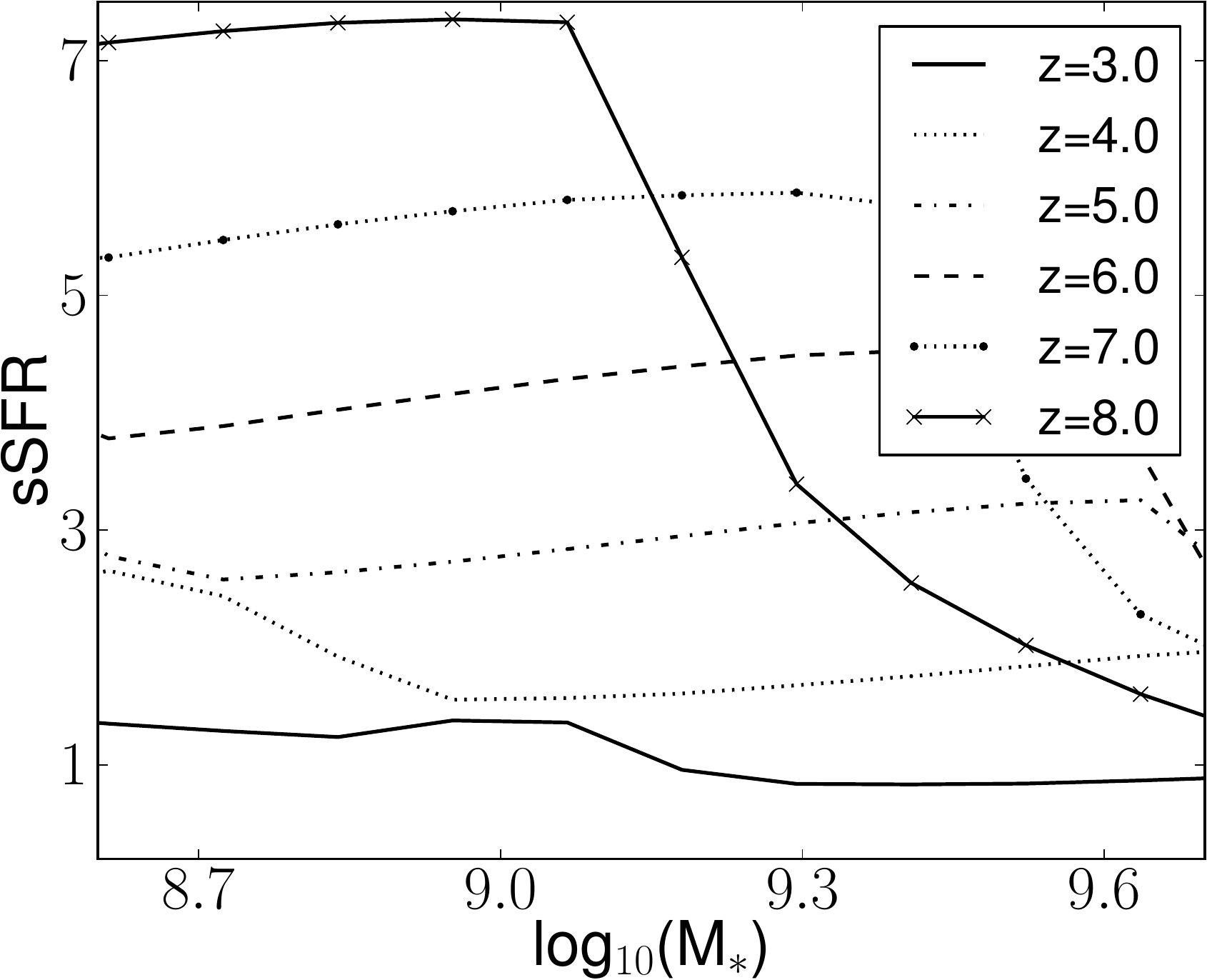} 
\caption{The mean sSFR computed using Eq.~(\ref{eqn:mean_ssfr}) as a function 
of $M_\ast$ at different redshifts.
}
\label{fig:mstar_ssfr}
\end{figure}

In our  model, even though the sSFR of galaxies forming at a given redshift 
is independent of the halo mass $M$ (see Eq.~(\ref{eqn:ssfr})), 
the total $M_\ast$ in 
them can differ, depending on the mass of the hosting dark matter halo. 
Conversely, galaxies of different sSFR (forming at different redshifts), 
can have the same $M_\ast$ depending on their dark matter halo mass. 
The average sSFR at redshift $z$ of galaxies of a given $M_\ast$ 
is given by
\be
sSFR(M_\ast,z) = \df{ \int~dz_c~sSFR(M,z_c)~\df{dn(M,z_c)}{dz_c}~\df{dM}{dM_\ast} }
                 { \int~dz_c~\df{dn(M,z_c)}{dz_c}~\df{dM}{dM_\ast} }, 
\label{eqn:mean_ssfr}
\ee
where halo mass $M = M(M_\ast,z_c)$ is related to the $M_\ast$ 
by Eq.~\ref{eqn:mstar} and sSFR(M,z,$z_c$) is given by Eq.~(\ref{eqn:ssfr}). 
In Fig.~\ref{fig:mstar_ssfr} we show the mean sSFR of galaxies as a function 
of their $M_\ast$ for  $3 \leq z \leq 8$. One can clearly see from the figure that 
for $5\times10^8 \leq M_\ast/M_\odot \leq 5\times10^{9}$ the sSFR is nearly constant 
for redshift range $z=3-7$. 
Similar trend has been noted at $z=2.3$ by the observational studies of 
\cite{reddy_pettini_12} as well. For highest redshift 
galaxies (i.e $z \sim 8$), our models predict a sharp decrease in the mean sSFR 
for $M_\ast \geq 10^9 M_\odot$.

\begin{figure}
\includegraphics[trim=0cm 5.0cm 9cm 5.0cm, clip=true, width =7.5cm, height=11.5cm, angle=0]
{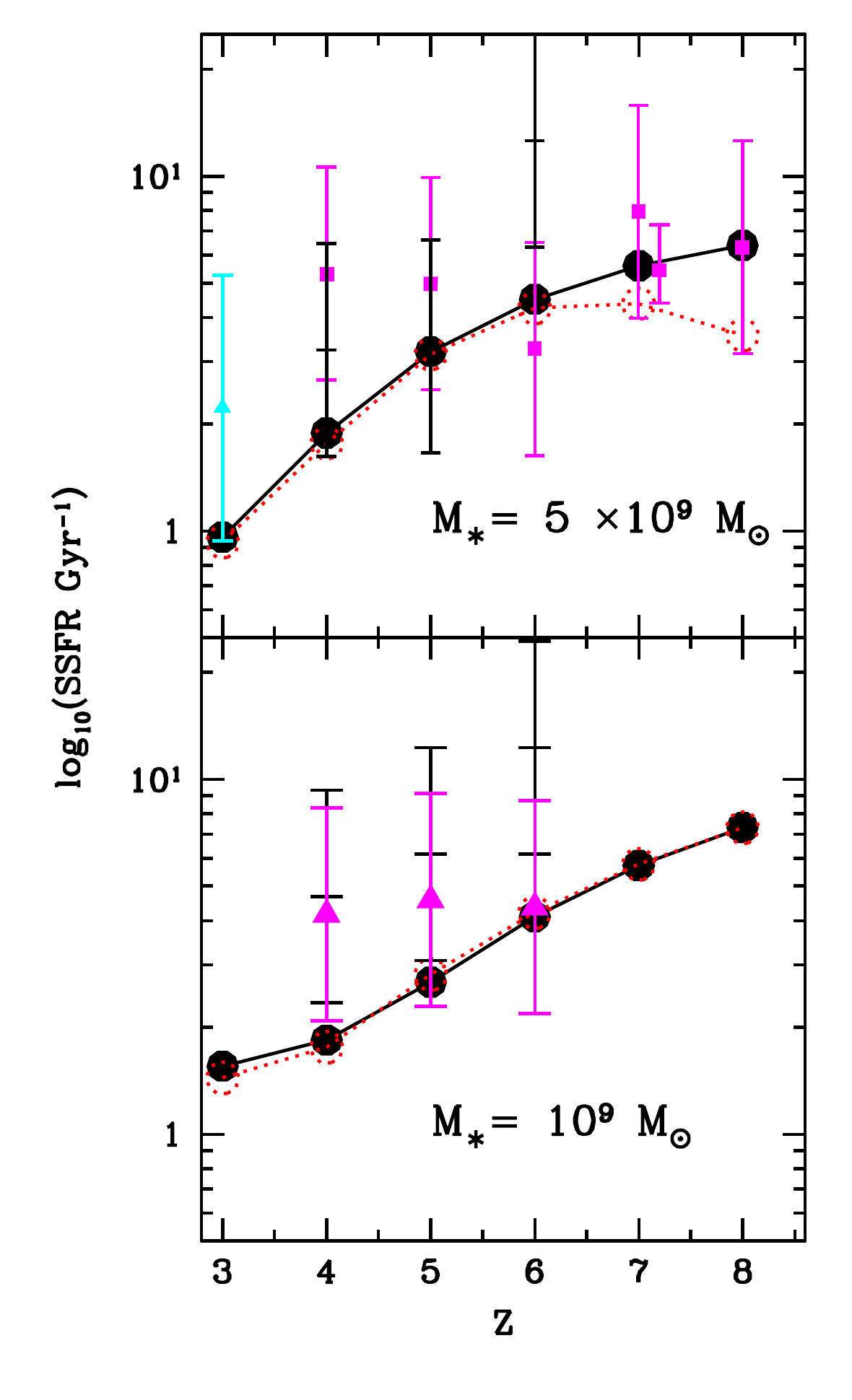} 
\caption{Top panel: The mean sSFR at $M_\ast = 5 \times 10^9 M_\odot$ as 
predicted by our models as a function of redshift. 
The data points and error bars are from 
(i) \protect \cite{feulner_05} ($z=3$, yellow-line)
(ii) \protect \cite{reddy_pettini_12} ($z=3$, green, triangle)
(iii) \protect \cite{stark_ellis_09_z46}($z=4-6$, cyan, hexagon)
(iv) \protect \cite{bouwens_illingworth_12} ($z=4-8$, magenta, square) and 
(v) \protect \cite{gonzalez_bouwens_12} ($z=4-8$, black, line).
Bottom panel: The mean sSFR at $M_\ast = 10^9 M_\odot$. 
The observed data are given by \protect \cite{gonzalez_bouwens_12} 
for rising (black, line) and constant (magenta, triangle) star formation models.   
In both panels, the solid black-curves are predictions of model A with luminosity 
dependent $\eta$ whereas the red-dotted lines are the predictions of model 
B with luminosity independent $\eta$. 
}
\label{fig:mstar_ssfr_m95}
\end{figure}

One particular quantity of interest is the sSFR at a fixed $M_\ast$. 
We have over plotted mean sSFR in Fig.~\ref{fig:mstar_ssfr_m95} 
at $M_\ast = 5 \times 10^9 M_\odot$ and $10^9 M_\odot$ 
as a function of redshift along with the observationally derived sSFR 
by various authors. 
The black solid line is the  
prediction of our model A whereas the red dotted line is for model B. 
These curves are obtained after averaging the 
sSFR of galaxies in mass bins centered around the specified $M_\ast$ and of  
width 0.3 dex. 
(See figure description for details about observed data).
Since sSFR is not a strong function of $M_\ast$ for 
$5 \times 10^8 \leq M_\ast/M_\odot \leq 5 \times  10^{9}$, we expect 
that the uncertainties in $M_\ast$ from SED fitting analysis will not strongly 
affect the mean sSFR at $M_\ast = 10^9 M_\odot$. 
Thus mean sSFRs of galaxies at this mass scale 
($10^9 M_\odot$) is a more robust quantity than stellar mass functions 
for constraining galaxy formation physics. 
We have also tabulated these mean sSFR in Table~\ref{tab1}.

The evolution of sSFR with redshift has been studied extensively in the literature, 
both observationally \citep{feulner_05, reddy_pettini_12, stark_ellis_09_z46, 
bouwens_illingworth_12, gonzalez_bouwens_12} and theoretically 
\citep{bouche_dekel_10, weinmann_neistein_11,krumholz_dekel_12,behroozi_13A}. 
Many of the theoretical studies suggest a rapid evolution of sSFR with redshift 
as $(1+z)^{2.4}$ \citep{bouche_dekel_10,weinmann_neistein_11,krumholz_dekel_12}.   
In contrast, earlier observational evidence suggested a nearly flat 
plateau for the sSFR for $z>3$ \citep{stark_ellis_09_z46,gonzalez_labbe_10}.
However most recent studies, including the effects of luminosity dependent 
dust obscuration and nebular emission in the SED fitting showed that 
sSFR does evolve with redshift \citep{bouwens_illingworth_12,gonzalez_bouwens_12}. 
In the case of model A incorporating the luminosity dependent dust 
extinction mean sSFR at $M_\ast = 5 \times 10^9 M_\odot$ evolves  
as $(1+z)^{2.4}$ over the redshift range $3 \leq z \leq 8$ 
which is in agreement with most of the theoretical studies. 
In our model B with luminosity independent $\eta$, the mean sSFR increases 
as $(1+z)^{1.9}$ for $3 \leq z \leq 7$. 
In general, sSFR at $M_\ast = 5 \times 10^9 M_\odot$ as predicted by our models 
compares well with the observationally derived sSFR within $1\sigma$ limits. 
For galaxies with $M_\ast = 10^9 M_\odot$, both our models A 
and B predict the evolution of sSFR as $(1+z)^2$. 

\section{Conclusions}
We have extended our physically motivated model to understand  
the evolution of $M_\ast$ - $M_{AB}$ scaling relations, stellar mass functions 
and sSFR of LBGs along with their UV luminosity functions in the redshift 
range $3 \leq z \leq 8$. 
In our model, galaxies are assumed to be formed in dark matter halos by 
continuous star formation which is proportional to the halo mass and 
extending over a period of a few dynamical time scales of the halo. 
This SFR combined with the halo formation rate gives the UV LF of LBGs which is 
compared with observed UV LF of LBGs to constrain the parameters 
related to the star formation. 

Using the observationally derived dust extinction parameter, $\eta$, 
we have obtained an estimate on $f_\ast$, the fraction of baryons being 
converted into stars over the lifetime of the galaxy. 
We found that $f_\ast$ is $\sim 0.2$ to $0.4$ for LBGs in the redshift range 
8 to 3, with major evolution occurring between redshifts 3 and 4. 
On the other hand, $f_\ast/t_{d}$, the parameter that scales as the SFR in LBGs 
is relatively constant and of order $\sim 0.2$ for $3 \leq z \leq 8$. 
This important result implies that the rate of conversion of baryons into stars 
is fairly constant for LBGs in the redshift range $3 \leq z \leq 8$. 
Therefore, during this period, the growth of the global star formation rate density 
can be explained solely by the evolution of dark matter halo mass function.

Using the value of $f_\ast$ determined by fitting the observed UV LF of LBGs, 
we obtained $M_\ast$ of individual halos at any 
given redshift. This enables us to further compute the correlation between 
$M_\ast$ and UV luminosity of LBGs,  their stellar mass functions and 
sSFR. 

We note that our model predictions of  $M_\ast$ - $M_{AB}$ correlations of LBGs at 
$z=3-5$ compare reasonably well with the observed data of \cite{reddy_pettini_12} 
and \cite{lee_ferguson_12}. 
Our models predict a slope of $\sim -0.40$ for the $M_\ast$ - $M_{AB}$ relation which 
is in good agreement with the bright end slope of this relation derived by 
\cite{lee_ferguson_12} using SED fitting. 
Further, normalization of the $M_\ast-M_{AB}$ relation in our models is found to decrease 
by a factor $3$ in the redshift range $4-7$ which is consistent 
with the findings of \cite{stark_schenker_13}. 

The stellar mass functions computed using our models also compares reasonably well 
with observed data for $M_\ast \leq 10^{10} M_\odot$ at $z=3$ and 
for $M_\ast \leq 10^{9} M_\odot$ at $z=4-6$. 
For higher $M_\ast$ our models underpredict stellar mass functions. 
However, we have shown that this discrepancy could be due to the 
uncertainties in the $M_\ast$ derived using SED fitting technique. 
\cite{mitchell_2013} showed that assumptions about dust, metallicity etc., will result in a 
scatter in the the observationally derived $M_\ast$ around it's true value. 
Further the uncertainty in the derived $M_\ast$ increases with increase in the true 
$M_\ast$ of the galaxy. 
We found that applying such a scatter in our theoretical models will up scatter many 
of the low stellar mass galaxies to the high mass end, significantly improving the agreement 
of our model predictions with observed data. 
Thus we conclude that, the systematic uncertainties in 
the derived $M_\ast$ should be accounted for 
while comparing theoretical models of stellar mass functions to observed data to 
constrain galaxy formation physics. 

We found that the sSFR of galaxies in our models at a given redshift is not a strong function 
of their $M_\ast$ for $5 \times 10^8 \leq M_\ast/M_\odot \leq 5 \times  10^{9}$, 
thus making it a much more robust quantity to constrain the physics of 
galaxy formation as compared to stellar mass functions. 
The sSFR computed using our models at $M_\ast$ of $5 \times 10^{9} M_\odot$ and 
$ 10^{9} M_\odot$  compares well with the observed data. 
Our models with luminosity dependent dust extinction predict that sSFR 
at $5 \times 10^{9} M_\odot$ evolves as $(1+z)^{2.4}$ for $3 \leq z \leq 8$ 
which is in agreement with previous theoretical 
studies. At $10^{9} M_\odot$ the sSFR  increases as $(1+z)^{2}$. 
In general our findings are  consistent the sSFR derived by 
\citep{bouwens_illingworth_12,gonzalez_bouwens_12} using SED fittings within $1\sigma$ limits
event hough the authors find only a moderate evolution in the mean sSFR with redshift.

In summary our physically motivated model presented in this paper explains 
various observables related to LBGs derived using SED fitting analysis along with their 
UV LFs. Remarkably, the key ingredient of this model, SFR in LBGs, extending for a dynamical 
time scale of the hosting dark matter halos shows no significant evolution for $3 \leq z \leq 8$. 
This evidently suggest that, the build up dark matter halos over the first two gigayears of the 
cosmic expansion, without invoking any redshift dependent efficiency of star formation activity in them, 
is sufficient to explain key observables related to LBGs.

Possible improvements to our work include exploring other forms of star formation histories
in individual halos, and/or using merger trees instead of the halo formation rate prescription
that we have used. However this will entail reproducing all the observables including the UV LFs 
and clustering of LBGs and LAEs, that our present model successfully explains.

\section*{Acknowledgments}
We thank Peter Behroozi for providing the observed data of 
stellar mass functions and sSFR. 
CJ acknowledges support from CSIR. 
CJ acknowledges support from CSIR. We also thank the referee for the 
useful comments that helped us to improve the paper.   

\def\aj{AJ}%
\def\actaa{Acta Astron.}%
\def\araa{ARA\&A}%
\def\apj{ApJ}%
\def\apjl{ApJ}%
\def\apjs{ApJS}%
\def\ao{Appl.~Opt.}%
\def\apss{Ap\&SS}%
\def\aap{A\&A}%
\def\aapr{A\&A~Rev.}%
\def\aaps{A\&AS}%
\def\azh{AZh}%
\def\baas{BAAS}%
\def\bac{Bull. astr. Inst. Czechosl.}%
\def\caa{Chinese Astron. Astrophys.}%
\def\cjaa{Chinese J. Astron. Astrophys.}%
\def\icarus{Icarus}%
\def\jcap{J. Cosmology Astropart. Phys.}%
\def\jrasc{JRASC}%
\def\mnras{MNRAS}%
\def\memras{MmRAS}%
\def\na{New A}%
\def\nar{New A Rev.}%
\def\pasa{PASA}%
\def\pra{Phys.~Rev.~A}%
\def\prb{Phys.~Rev.~B}%
\def\prc{Phys.~Rev.~C}%
\def\prd{Phys.~Rev.~D}%
\def\pre{Phys.~Rev.~E}%
\def\prl{Phys.~Rev.~Lett.}%
\def\pasp{PASP}%
\def\pasj{PASJ}%
\def\qjras{QJRAS}
\def\rmxaa{Rev. Mexicana Astron. Astrofis.}%
\def\skytel{S\&T}%
\def\solphys{Sol.~Phys.}%
\def\sovast{Soviet~Ast.}%
\def\ssr{Space~Sci.~Rev.}%
\def\zap{ZAp}%
\def\nat{Nature}%
\def\iaucirc{IAU~Circ.}%
\def\aplett{Astrophys.~Lett.}%
\def\apspr{Astrophys.~Space~Phys.~Res.}%
\def\bain{Bull.~Astron.~Inst.~Netherlands}%
\def\fcp{Fund.~Cosmic~Phys.}%
\def\gca{Geochim.~Cosmochim.~Acta}%
\def\grl{Geophys.~Res.~Lett.}%
\def\jcp{J.~Chem.~Phys.}%
\def\jgr{J.~Geophys.~Res.}%
\def\jqsrt{J.~Quant.~Spec.~Radiat.~Transf.}%
\def\memsai{Mem.~Soc.~Astron.~Italiana}%
\def\nphysa{Nucl.~Phys.~A}%
\def\physrep{Phys.~Rep.}%
\def\physscr{Phys.~Scr}%
\def\planss{Planet.~Space~Sci.}%
\def\procspie{Proc.~SPIED}%
\let\astap=\aap
\let\apjlett=\apjl
\let\apjsupp=\apjs
\let\applopt=\ao

\bibliographystyle{mn2e}	
\bibliography{ref.bib,sfr.bib,ref-1.bib}		

\end{document}